# Percolation Model of Insider Threats to Assess the Optimum Number of Rules

Jeremy Kepner, Vijay Gadepally, Pete Michaleas
MIT Lincoln Laboratory, Lexington, MA, U.S.A.

*Abstract*— Rules, regulations, and policies are the basis of civilized society and are used to coordinate the activities of individuals who have a variety of goals and purposes. History has taught that over-regulation (too many rules) makes it difficult to compete and under-regulation (too few rules) can lead to crisis. This implies an optimal number of rules that avoids these two extremes. Rules create boundaries that define the latitude an individual has to perform their activities. This paper creates a Toy Model of a work environment and examines it with respect to the latitude provided to a normal individual and the latitude provided to an insider threat. Simulations with the Toy Model illustrate four regimes with respect to an insider threat: under-regulated, possibly optimal, tipping-point, and over-regulated. These regimes depend up the number of rules (N) and the minimum latitude ($L_{min}$) required by a normal individual to carry out their activities. The Toy Model is then mapped onto the standard 1D Percolation Model from theoretical physics and the same behavior is observed. This allows the Toy Model to be generalized to a wide array of more complex models that have been well studied by the theoretical physics community and also show the same behavior. Finally, by estimating N and $L_{min}$ it should be possible to determine the regime of any particular environment.

I. INTRODUCTION

Rules, regulations, and policies coordinate the activities of individuals who have a variety of goals and purposes [Brennan & Buchanan 1988] and are the basis of civilized society. Rules are the primary tool for creating safe, secure, and fair environments for individuals, organizations, and communities [Schauer 1991]. Rules work by establishing acceptable boundaries for activities. The space between the boundaries is the latitude individuals have to perform activities. Over-regulation (too many rules) provides insufficient latitude for individuals to accomplish their activities. "History has taught us that over-regulated economies have difficulty competing" [Shore 1998]. Under-regulation (too few rules) is also bad and can to lead to crisis. In particular, "if the spell of no crisis is long enough, the regulation level may drop to zero, despite the fact that the socially optimal regulation level remains positive" [Aizenman 2009]. This implies an optimal number of rules that avoids these two extremes [Barro 1986]. This paper seeks to develop models that qualitatively illustrate this behavior and can be used as a starting point for a more detailed, quantitative approach for determining the optimal number of rules in a particular environment.

Decision latitude is the ability to make work-related decisions [Karasek 1979]. Rules create boundaries that determine the decision latitude of the individual with respect to a particular activity. Low decision latitude can be associated with adverse health consequences, such as coronary heart disease [Kuper & Marmot 2003], due to increased workplace stress. Measuring decision latitude or the distance an individual action is from an acceptable boundary has been mostly a qualitative endeavor. Crowd-sourced ethics is one method to measure the distance to a boundary [Casey & Sheth 2013]. For example, asking individuals in a specific environment about the desirability (or undesirability) of a specific hypothetical action can be used to establish the quantitative distance that action is from an acceptable boundary.

Breaking a rule involves crossing a boundary established by the rule. Many boundary crossings are done to serve the larger goals of an organization [Umphress & Humphries 2011] or to maintain safety, order, or equality [Verkuyten et al 1994]. In a survey of over 2,000 executive assistants [Klieman 1996], it was found that:

- 10% destroyed or removed damaging information;
- 6.5% wrote documents with misleading or false information; and
- 5.1% falsified vouchers or expense accounts.

All of this activity was performed by employees to benefit their bosses and/or their organizations. Such organizationally motivated boundary crossings may be the more frequent class of boundary crossing and may be indicative of low decision latitude and perhaps over-regulation. For example, if a neighborhood has stop signs on every corner, it will increase the number of stop signs that are ignored. Frequent boundary crossings effectively remove those boundaries and create opportunities for insider threats to increase their latitude.

An insider can be defined as a current or former employee, contractor, or business partner who has or had authorized access to an organization. Likewise, an insider threat can be defined as an insider who intentionally exceeded or intentionally used their access in a manner that negatively affected the organization [Silowesh & Nicoll 2013]. This definition excludes accidents, incompetence, or whistle-blowers. For example, Chelsea Manning was intentionally trying to do harm to the United States and would fall well within the insider threat definition. Edward Snowden believes he is a whistle-blower while the United States Intelligence Community believes he is an insider threat.

One of the main goals of an insider threat is to increase their decision latitude. This can be accomplished using a standard OODA loop [Boyd 1996]:

**Observe** boundaries that are being crossed,
**Orient** themselves to cross those boundaries,
**Decide** to cross those boundaries, and
**Act** on the new latitude they have achieved.

Through this process an insider threat elevates their privileges; gains knowledge about control measures; and may be able to bypass security measures designed to prevent, detect, or react to unauthorized access [Myers et al 2009]. A typical example is an insider acquiring a co-workers password [Claycomb et al 2013]. In addition, an insider threat can be defined as an entity that violates a security policy using legitimate access [Bishop & Gates 2008]. In an over-regulated environment, frequent boundary crossing by normal individuals allows the insider to increase their decision latitude by observing and mimicking the same behavior.

Technological measures for addressing insider threats are an active area of research. One approach is better cryptographic solutions [Yakoubov et al 2014] that add limitation to the use of information. The costs of these approaches are a key aspect of their adoption and there is active research to develop technologies that minimize their limitations [Kepner et al 2014].

Given the importance and cost implementing of rules this work attempts to provide models for estimating the optimal number of rules in the context of an insider threat. This work follows a two-step approach. Step 1 is a simple Toy Model based on a 1D concept of latitude. The Toy Model allows simple estimates of the impact of rules on latitude and illustrates the key concepts of over-regulation, under-regulation, and a tipping-point between these two regimes. Step 2 changes the terminology of the Toy Model to correspond to a Percolation Model. Percolation is widely use model in theoretical physics for understanding a wide range of phenomena. The Percolation Model leverages over 100,000+ journal articles on percolation [Stuaffer & Aharony 1992], critical phenomena [Hohenberg & Halperin 1977], phase transitions [Sinai 1982], and the renormalization group [Wilson 1975, Binney et al 1992] and allows rapid generalization of the Toy Model to more complex models that have already been well studied. Finally, and perhaps most importantly, the behavior of the Percolation Model has been shown to be similar to that of many more complex models. Thus, the Toy Model is capturing the essence of the behavior found in these systems. Using a more complex model may create more quantitatively accurate models, but it is likely that qualitative behavior will remain the same.

## II. TOY MODEL

The Toy Model (see Figure 1) represents an rule based environment as a simple linear 1D domain of points X where $0 < X < 1$. Hard boundaries are placed at $X = 0$ and $X = 1$. A normal individual performs their required activities within this domain. In the initial configuration the number of rules is $N = 0$ and the average latitude of a normal individual in this environment is $L_{normal}(N=0) = 1$. As new rules are added, new boundaries are created at random at a new location in the domain. If there are N rules, there will be N internal boundaries. Sorting the boundary locations in increasing order results in a boundary at $X_1$, another boundary at $X_2$, such that

$$X_1 < X_2 < \ldots < X_i < \ldots X_N$$

Taking the difference between neighboring boundaries results in a sequence of N+1 latitudes

$L_1 = X_1$
$L_2 = X_2 - X_1$
…
$L_i = X_i - X_{i-1}$
…
$L_N = X_N - X_{N-1}$
$L_{N+1} = 1 - X_N$

The sum of the latitudes is always $\Sigma L_i = 1$. So the average latitude for the normal individual in the environment is always

$$L_{normal} = 1/(N + 1)$$

As more rules are added and more boundaries are created the average latitude of the normal individual decreases.

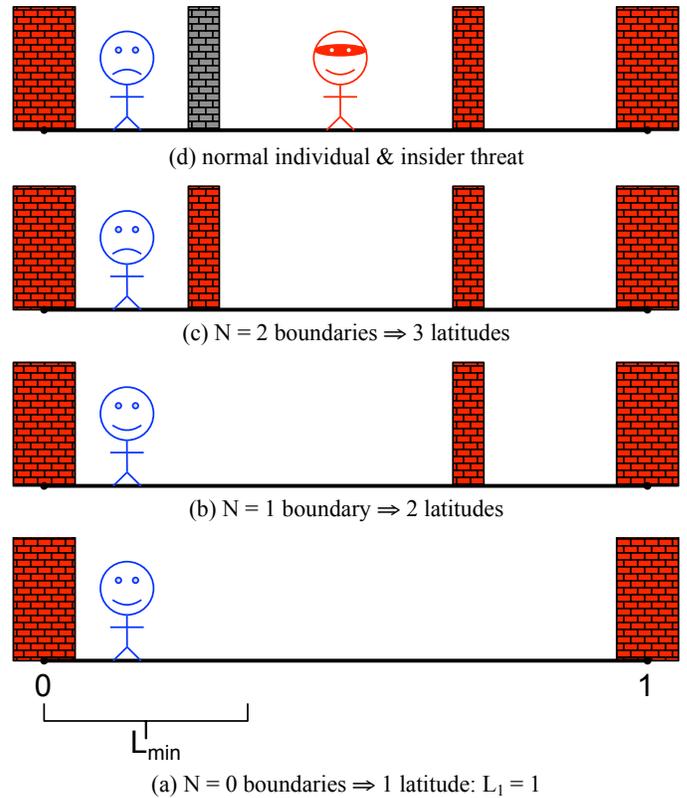

(d) normal individual & insider threat

(c) N = 2 boundaries ⇒ 3 latitudes

(b) N = 1 boundary ⇒ 2 latitudes

(a) N = 0 boundaries ⇒ 1 latitude: $L_1 = 1$

Figure 1. Toy Model of an environment with a normal individual and an insider threat. The environment has boundaries at coordinates 0 and 1. $L_{min}$ is the minimal latitude required for normal activity. From bottom to top: (a) no internal boundaries (N=0) implies a single latitude $L_1 = 1$; (b) one internal boundary (N=1) implies two latitudes; (c) two internal boundaries (N=2) implies three latitudes, one of which is less than $L_{min}$; (d) the latitude less than $L_{min}$ is eliminated for the adversary.

Let, $L_{min}$ denote the minimum latitude an individual needs to perform their required activities. If any two boundaries are less than $L_{min}$ apart, then one of the boundaries is likely to be crossed. The normal individual will mostly attempt to remain within these boundaries and so their average latitude is unchanged. However, because the boundary needs to be crossed often, it is effectively removed for the insider threat. Boundaries that are likely to be crossed increase the average latitude of the insider threat. The number of insider threat boundaries $N_{threat}$ is equal to the number of latitudes satisfying $L_i > L_{min}$. As before, all the threat latitudes must sum to 1 so that the average threat latitude is given by

$$L_{threat} = 1/(N_{threat} + 1)$$

A simulation of the Toy Model behavior is shown in Figure 2. The code for this simulation is provided in the Appendix. Figure 2 shows the average latitude versus the number of rules for normal individuals and insider threats. As expected, the insider threat latitude first decreases with the number rules, reaches a minimum, and then increases as the number of rules increases. The minimum threat latitude occurs at

$$N_{min} = 1/L_{min}$$

If $N > N_{min}$, then threat latitude increases.

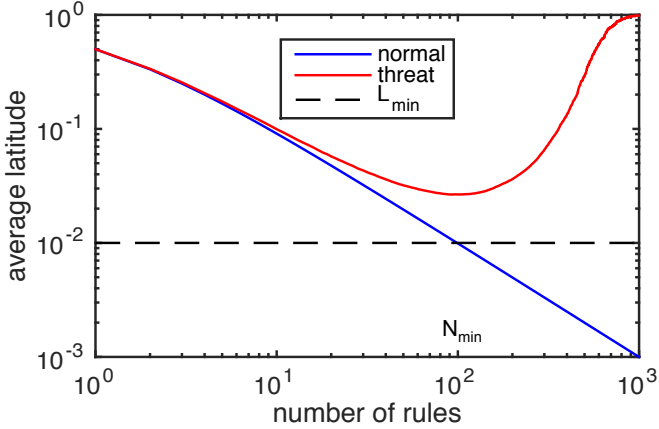

Figure 2. Toy Model average latitude versus number of rules for a normal individual and an insider threat averaged over 100 simulations. Insider threat latitude reaches a minimum when the number rules is $N_{min} = 1/L_{min}$.

The ratio of the threat latitude to the normal latitude is shown in Figure 3. For modest numbers of rules the threat latitude is comparable to the normal latitudes. As N approaches $N_{min}$ the ratio starts to increase. In this particular instance, the threat latitudes is ~3 times the normal latitude at $N=N_{min}$. The ratio increases dramatically as N surpasses $N_{min}$.

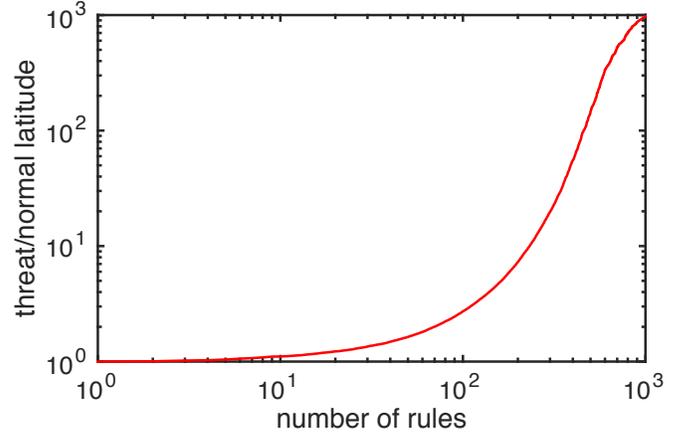

Figure 3. Toy Model ratio of relative latitude of insider threat to normal individual. For modest numbers of rules the threat latitude is comparable to the normal latitudes. As N approaches $N_{min}$ the ratio starts to increase. In this particular instance, the threat latitudes is ~3 times the normal latitude at $N=N_{min}$. The ratio increases dramatically as N surpasses $N_{min}$.

Figure 2 and Figure 3 illustrate four possible regimes that are described in Figure 4. The regimes are under-regulated, possibly optimal, tipping point, and over-regulated. In the under-regulated regime, the threat has a large latitude that is very similar to that of the normal individual. At the tipping-point, the threat latitude is minimized and the normal latitude is near the $L_{min}$. In the over-regulated regime, the threat latitude is increasing and the normal latitude is below $L_{min}$. We speculate that a possibly optimal regime exists between the under-regulated regime and the tipping-point.

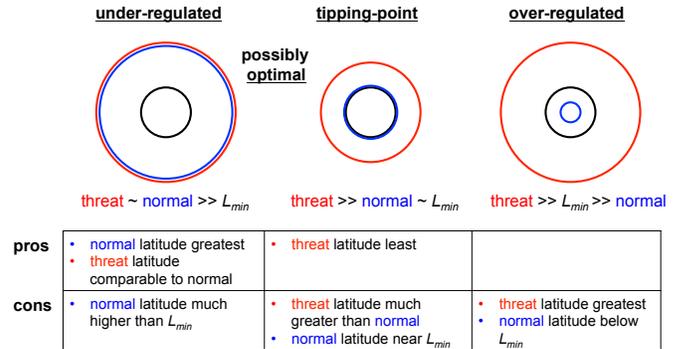

Figure 4. Toy Model regimes. In the under-regulated regime, the threat has a large latitude that very similar to that of the normal individual. At the tipping-point, the threat latitude is minimized and the normal latitude is near the $L_{min}$. In the over-regulated regime, the threat latitude is increasing and the normal latitude is below $L_{min}$. We speculate that a possibly optimal regime exists between the under-regulated regime and the tipping-point.

### III.  PERCOLATION MODEL

The Toy Model produces behaviors that are consistent with intuition: under-regulation, tipping-point, and over-regulation. However, it is not realistic to assume that complex human activities can be captured with such a simple model (at least not without more evidence). A more realistic model would likely have additional dimensions and more complex interactions among dimensions. Conducting a detailed analysis of the implications of adding dimensions and more complex interactions would be an overwhelming undertaking. Fortunately, the theoretical physics community has already made this investment. By changing the terminology of the Toy Model and converting it to a Percolation Model, it is possible to tap into this prior investment in theoretical work.

Theoretical physics has extensively explored models of complex systems that show behavior similar to the Toy Model. These models fall under a wide range of names that include percolation, critical phenomena, phase transitions, and the renormalization group. These models have also been applied to modeling forest fires, oil fields, diffusion, and a range of other phenomena [Stauffer & Aharony 1992].

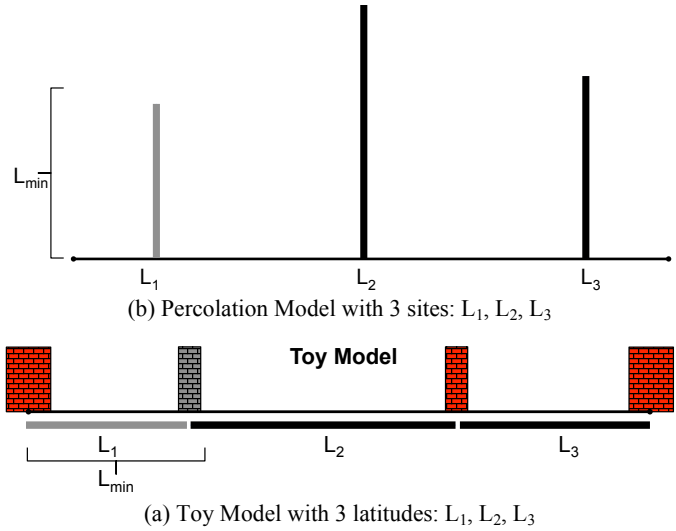

Figure 5. Toy Model and Percolation Model. (a) Toy Model with three latitudes where $L_1 < L_{min}$. (b) Percolation Model that is equivalent to the Toy Model in (a). Site 1 is said to be occupied because $L_1 < L_{min}$.

The simplest of these models is 1D percolation. The Toy Model can be transformed into the Percolation Model in two steps. First, the latitudes $L_i$ are "rotated" so that they are sitting on "sites" (see Figure 5) and when $L_i < L_{min}$ the site is said to be "occupied." Second, the latitudes are turned into probabilities that the site is occupied. The latitudes $L_i$ are sorted differences whose cumulative probability distribution is given by an exponential (see Figure 6)

$$P(N,L) = 1 - \exp[-(N+1) L]$$

This completes the transformation. Part of power of the Percolation Model is that it is effectively defined by only two parameters: the dumber of dimensions (in this case 1D) and the probability that a site is occupied (in this case $P(N,L)$).

This probability also provides a mechanism for computing the exact solution of the Toy Model since

$$N_{threat} = (1 - P(N,L_{min})) N$$

resulting in

$$L_{exact}(N,L_{min}) = 1/((1 - P(N,L_{min})) N + 1)$$
$$= 1/(N \exp[-(N+1) L_{min}] + 1)$$

The above expression is minimized when $N_{min}=1/L_{min}$ and results in

$$L_{exact}(N_{min},L_{min}) = 1/(N_{min} \exp[-(L_{min} + 1)] + 1)$$

which for $L_{min} \ll 1$ and $N_{min} \gg 1$ is

$$L_{exact}(N_{min},L_{min}) \approx e\, L_{min}$$

Transforming the Toy Model into the Percolation Model connects it with the extensive theoretical work on percolation. One of the primary areas of study in percolation theory is the probability of clusters of occupied sites (i.e., areas of large latitude). The average size of a cluster in an infinite site 1D Percolation Model is [Stauffer & Aharony 1991]

$$S(P) = (P_c + P)/(P_c - P)$$

where $P_c$ is the percolation threshold. For an infinite site 1D percolation $P_c = 1$. In the Percolation Model of an insider threat the average latitude of a site is

$$L_{avg} = 1/(N+1)$$

Combining the formulas for $L_{avg}$ and $S(P)$ gives the threat latitude in an infinite site 1D Percolation Model

$$L_{threat} = L_{avg}\, S(P(N,L_{min}))$$

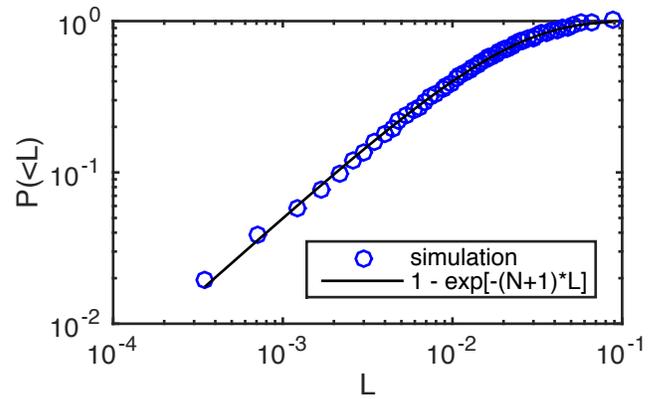

Figure 6. Probability of a percolation site being occupied versus latitude. Latitudes are sorted differences that are well modeled by an exponential distribution.

Figure 7 compares the Toy Model value for threat latitude with the equivalent value computed using the Percolation. The infinite site 1D Percolation Model is qualitatively equivalent to the Toy Model simulations. Small differences between the models are due to finite site simulations versus an infinite site model. Finite percolation models can also be calculated analytically, but they are beyond the scope of this work.

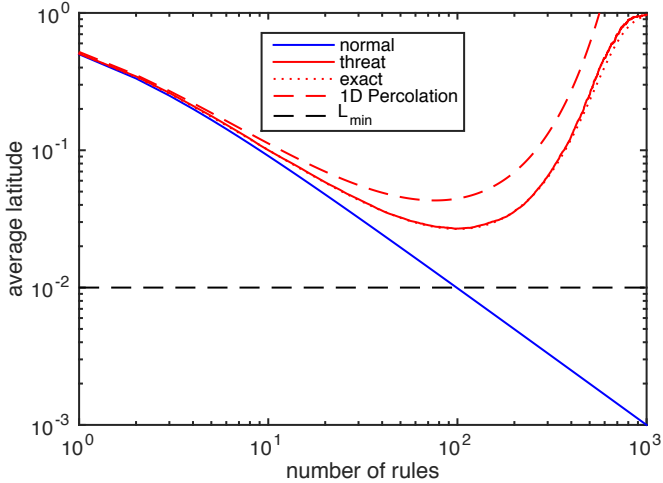

Figure 7. Toy Model simulations of a normal individual and an insider threat along with the exact solution to the Toy Model and the infinite site Percolation Model. All models are in qualitative agreement. Differences are the result of the Toy Model being finite and Percolation Model being infinite.

## IV. IMPLICATIONS

The qualitative agreement between the Toy Model and the Percolation Model allows the Toy Model to be connected with far more complex models. Percolation has been studied in many systems (see Figure 8) and all exhibit the same qualitative behavior that is determined by a single parameter P. This suggests the qualitative behavior of the 1D insider threat model may also be similar to higher fidelity insider threat models. For example, in more complex percolation models the percolation threshold $P_c$ gets smaller as the model gets more complex (see Figure 8). This would mean that $N_{min}$ becomes smaller in more complex systems and that insider threat latitude starts increasing with a smaller number of rules.

## V. SUMMARY & FUTURE WORK

Rules are the primary means of coordinating the activities of individuals who have a variety of goals and purposes. Over-regulation and under-regulation can both be exploited by insider threats. The Toy Model illustrates this phenomena quantitatively. Adapting the Toy Model to the Percolation Model confirms this phenomena over a wide range models and connects its with a vast literature that has been successfully applied to many fields.

This work suggest that collecting specific pieces of information about the rules in an environment could provide incite into whether an environment is under-regulated, near the tipping-point, or over-regulated. In particular, estimating the number of rules, boundary crossings, and $L_{min}$ could be use to identify the regime. Possible approaches to collecting this information could be:

- Counting the number of rules [Antoniou et al 1999],
- Measuring the volume of materials used to describe the rules,
- Examining the number of exception requests,
- Examining the number accidental boundary crossing, and
- Surveying the individuals in the environment about crossings.

This work provides a rich theoretical framework for incorporating and understanding such data and increases the value of collecting such data.

| Dimension | Lattice | Site Neighbors | $P_c$ |
|---|---|---|---|
| 1d | linear | 2 | 1 |
| 2d | honeycomb | 3 | 0.696 |
| 2d | square | 4 | 0.593 |
| 2d | triangular | 6 | 1/2 |
| 3d | diamond | 4 | 0.430 |
| 3d | dimple cubic | 6 | 0.312 |
| 3d | body centered cubic | 8 | 0.246 |
| 3d | face centered cubic | 12 | 0.198 |
| 4d | hypercubic | 8 | 0.197 |
| 5d | hypercubic | 10 | 0.141 |
| 6d | hypercubic | 12 | 0.107 |
| 7d | hypercubic | 14 | 0.089 |
| | Bethe lattice (Cayley graph) | z | 1/(z-1) |

Figure 8. Percolation has been studied in a wide range of systesms. The table shows a select few with varying number of dimensions, geometry and numbers of neighbors. $P_c$ is the probability at which the system is likely to become fully connected.

APPENDIX: MODEL CODE

The Toy Model and the Percolation Model can be computed using the following Matlab (or GNU Octave) code.

```
Ntrial = 100;       % Number simulations to average
Nrule = 1000;       % Maximum number of rules to test
Lmin = 0.01;        % Set the minimum latitude
Lnormal = zeros(1,Nrule);    % Initialize Lnormal
Lthreat = zeros(1,Nrule);    % Initialize Lthreat

% TOY MODEL
for itrial = 1:Ntrial       % Loop over each trial
  disp(num2str(itrial));    % Print trial number
  X = rand(1,Nrule);        % Create boundaries
  for irule = 1:Nrule       % Loop over rules
    % Add 0 and 1 and to boundaries sort
    Xnormal = sort([0 X(1:irule) 1]);
    % Compute the normal latitudes
    Ln = diff(Xnormal);
    % Average and tally trial
    Lnormal(irule) = Lnormal(irule) + mean(Ln);
    % Eliminate boundaries below Lmin
    Xthreat = unique([0 Xnormal(find(Ln>Lmin)+1) 1]);
    % Compute the threat latitudes
    Lt = diff(Xthreat);
    % Average and tally trial
    Lthreat(irule) =  Lthreat(irule) + mean(Lt);
  end                       % End loop over rules
end                         % End loop over trials
Lnormal = Lnormal ./ Ntrial; % Avg over trial
Lthreat = Lthreat ./ Ntrial; % Avg over trial

% PERCOLATION MODEL
N = 1:Nrule;                % Number  of sites
Pmin = 1 - exp(-(N+1)*Lmin); % Site probability
SPmin = (1+Pmin)./(1-Pmin);  % Avg cluster size
Lavg = 1./(N+1);             % Avg site latitude
Lexact = 1./((1-Pmin).*N + 1);   % Exact answer

% PLOT RESULTS
figure;
loglog(N,Lnormal,'b',N,Lthreat,'r',N,Lexact,'r:',N,LthreatPerc,'r--',[1 Nrule],[Lmin Lmin],'k--');
xlabel('number of rules'); ylabel('average latitude');  legend('normal','threat','exact','1D Percolation','L_{min}','Location','North');
axis([1 Nrule 1/Nrule 1]);
```